\documentclass[fleqn,10pt]{wlscirep}
\usepackage[utf8]{inputenc}
\usepackage[T1]{fontenc}
\usepackage{makecell}
\usepackage{quantikz}
\usepackage{svg}
\title{Structured leakage in OAM-encoded qubits revealed by distributed quantum feature extraction}

\author[1]{Yayu Mo}
\author[2]{Nazifa Rumman}
\author[1]{Sanjeev Shapkota}
\author[2]{Thomas A. Searles}
\author[1,*]{Sanjaya Lohani}
\affil[1]{Department of Electrical and Computer Engineering, Southern Methodist University, Dallas, 75205, USA}
\affil[2]{Department of Electrical and Computer Engineering, University of Illinois Chicago, Chicago, IL, 60607, USA}

\affil[*]{Corresponding author: slohani@smu.edu}

\begin{abstract}

Orbital angular momentum (OAM) modes have emerged as an important platform for high-dimensional quantum communication but remain vulnerable to environmental noise and channel imperfections. Here we introduce a logical-qubit representation of OAM propagation in which modes evolve as qubits through noisy quantum channels. In this representation, turbulence manifests as leakage from the logical OAM subspace into turbulence-induced leakage components, providing a unified description of degradation in both single-photon and coherent-beam regimes. We show that turbulence-induced leakage exhibits systematic dependence on turbulence strength and OAM order, revealing structured signatures of information redistribution beyond the logical subspace. Leveraging this insight, we develop a distributed quantum feature-extraction scheme that encodes high-dimensional OAM states into quantum representations and employs quantum convolutional operations to extract turbulence-robust features. Simulations and experimental measurements demonstrate robust extraction of these features across a broad range of turbulence strengths, OAM orders, and propagation conditions. Our results establish a direct link between OAM-channel dynamics, structured leakage, and quantum learning, providing a physics-informed framework for characterizing turbulence-degraded high-dimensional optical states and extracting robust signatures of turbulence-induced information redistribution.

\end{abstract}
\begin{document}

\flushbottom
\maketitle

\section{Introduction}

High-dimensional quantum communication exploits the rich structure of photonic degrees of freedom to increase information capacity beyond conventional qubit-based encoding schemes. Among these degrees of freedom, orbital angular momentum (OAM) modes are particularly attractive because they provide an effectively unbounded Hilbert space for quantum information encoding and have become a promising platform for free-space quantum communication \cite{OAMComm1, OAMComm2, OAMComm3, OAMComm4}, quantum networking \cite{OAMQnet1, OAMQnet2, lin2025quantum}, and high-dimensional quantum information processing \cite{OAMQinfo1, OAMQinfo2, lin2025two, zheng2026grover}. However, practical deployment of OAM-based communication remains limited by environmental noise, particularly atmospheric turbulence, which induces phase distortions, mode mixing , and crosstalk that degrade coherence and transmission fidelity \cite{turb_OAM_mux, OAM_turb3} in both single-photon and coherent-beam regimes \cite{turb_oam_single_photon, coherent_beam_turb}. Recent machine-learning approaches have demonstrated impressive performance in classifying, reconstructing, and denoising structured optical fields under turbulence \cite{lohani2018turbulence, ml_oam_spec_turb}, yet they largely treat degradation as a signal-processing problem and provide limited insight into the underlying quantum-state dynamics. Consequently, a fundamental question remains unresolved: how do noisy quantum channels reshape OAM-encoded information and can the resulting channel-induced degradation reveal physically meaningful signatures that remain accessible beyond the ideal logical subspace?

Experimental studies have investigated the transmission of OAM-based quantum states through atmospheric turbulence, including demonstrations of OAM-entanglement distribution and BB84 quantum key distribution under realistic channel conditions \cite{entanglement_turb, BB84_turb, BB84_turb2}. These works have demonstrated the feasibility of employing OAM states for quantum information transmission while highlighting the influence of turbulence on quantum-state fidelity and communication performance. Beyond communication-oriented studies, Sorelli et al. \cite{unitary_turb_relation} examined turbulence-induced evolution within OAM quantum subspaces and interpreted the associated fluctuations through effective unitary transformations. More recently, Maqsood et al. \cite{simulation_quantum_gates_oam_qubit} demonstrated that OAM-encoded qubits can support the simulation of quantum-gate operations, further highlighting the utility of OAM modes as carriers of quantum information. Together, these studies establish the viability of OAM states for quantum information processing and communication. However, turbulence is typically characterized by its impact on state fidelity, crosstalk, or communication performance, leaving the underlying quantum-channel dynamics largely unexplored. In particular, it remains unclear how turbulence redistributes information within and beyond an encoded OAM logical subspace, and whether the resulting degradation generates structured signatures that can be exploited for characterization and robust feature extraction.

Here, we present a logical-qubit framework for OAM propagation in simulations and experiments and show that turbulence-induced degradation manifests as structured leakage, namely the systematic redistribution of probability amplitude beyond an encoded OAM subspace. Unlike random mode mixing, the resulting leakage patterns exhibit clear dependence on turbulence strength, propagation conditions, and OAM order. This perspective is motivated by the observation that atmospheric turbulence redistributes information among OAM modes rather than simply destroying it \cite{influence_turb,bart2025classification,Rumman:26}. To capture this behavior, we model OAM-state evolution using noisy quantum channels that mimic realistic open-system dynamics and include both turbulence-derived and conventional quantum-noise processes. Within this framework, turbulence-induced crosstalk appears as leakage from the encoded subspace into turbulence-scattered components, producing effective non-trace-preserving behavior when the dynamics are viewed within a truncated OAM basis. This channel-theoretic description suggests that turbulence does not merely reduce fidelity but redistributes probability amplitude in a structured manner, giving rise to physically meaningful signatures that may be exploited for characterization and feature extraction.

Motivated by the existence of these structured leakage signatures, we develop a distributed quantum feature-extraction framework for analyzing turbulence-distorted OAM intensity distributions. A central challenge is the efficient encoding of high-dimensional OAM intensity patterns into quantum representations without excessive quantum-resource overhead or additional encoding noise. To address this challenge, measured OAM intensity distributions are partitioned into smaller image patches and encoded into distributed quantum subsystems, enabling scalable quantum-state preparation while reducing resource requirements and mitigating encoding noise \cite{yayu_2025qce_poster_distri_comp,yayu_2026dcas_distri_img_encoding}. This distributed architecture is particularly attractive for high-dimensional OAM data, where direct quantum encoding of full-resolution intensity images can become resource intensive. While classical convolutional neural networks provide an effective framework for analyzing turbulence-distorted optical images\cite{cnn_oam_detection, lohani2020generative, cnn_oam_detection2}, our focus is not computational advantage but whether leakage-related signatures remain accessible within quantum representations that naturally interface with quantum-information-processing workflows. By preserving localized spatial structures while reducing encoding complexity, the distributed framework facilitates the analysis of leakage-related signatures that emerge in turbulence-distorted OAM patterns. Inspired by convolutional feature extraction in classical learning architectures \cite{conv_feature1,conv_feature2}, we employ quantum convolutional operations to identify features associated with turbulence-induced information redistribution. Through simulations and experimental measurements, we find that signatures associated with structured leakage persist across a range of turbulence strengths, OAM orders, and propagation conditions.

In this work, we define channel dynamics as the turbulence-induced evolution of OAM information, as characterized by mode-coupling scattering matrices, effective channel operators, and leakage beyond the encoded logical subspace. The central contribution of this paper is a quantitative framework that connects channel dynamics, structured leakage, and high-dimensional feature representations. By jointly analyzing channel evolution, leakage behavior, and extracted features, we examine how turbulence-induced degradation manifests itself in both intensity and feature spaces, providing a framework for characterizing information redistribution in high-dimensional OAM systems.

\begin{figure*}[h]
    \centering
    \includegraphics[width=0.85\linewidth]{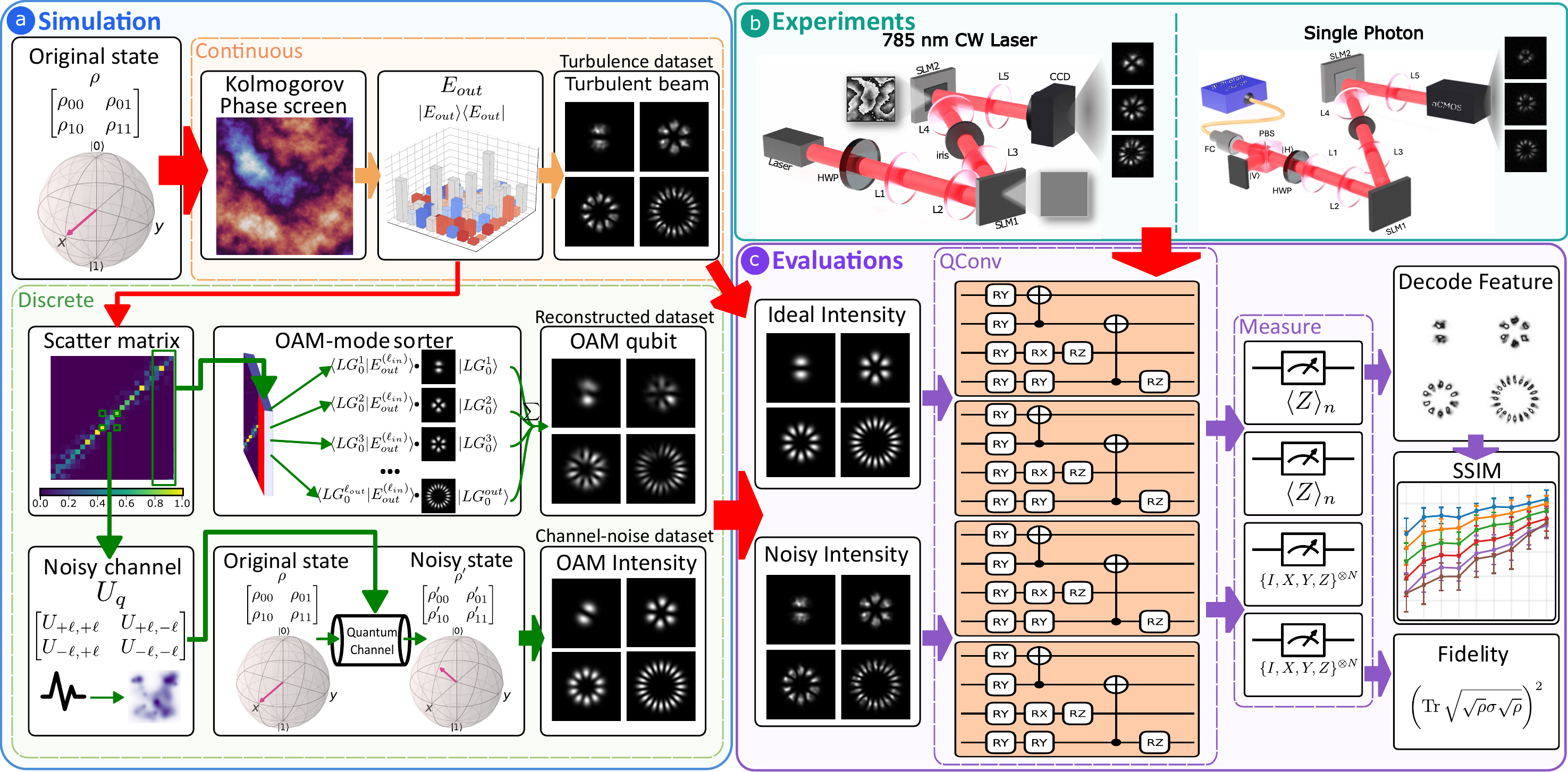}
    \caption{Main pipeline of our work. (a) The simulation block consists of three sections: Original State, Continuous, and Discrete. (b) The experiments block contains 785 nm CW Laser and single-photon experiment setup schematics. (c) The evaluation block contains the quantum convolutional feature extraction operation (QConv), measurement sections (decoded expectation for calculating structural similarity index measurement and ideal state tomography for fidelity).}
    \label{fig: intro_fig}
\end{figure*}

\section{Results}
\subsection{Model}
Fig.~\ref{fig: intro_fig} illustrates the complete workflow of our framework, which integrates OAM-qubit channel modeling, experimental validation, and distributed quantum feature extraction. We begin by representing OAM modes as logical qubits and studying their evolution through turbulence and noisy quantum channels in both continuous and discrete settings. This formulation enables us to connect physical propagation, quantum-channel dynamics, and feature-space representations within a unified framework.

We first construct simulation datasets in both continuous and discrete representations. In the continuous setting, we propagate an initial quantum state $\rho$ through a Kolmogorov phase screen using a continuous-wave (CW) laser model to generate the turbulent optical field $E_{\mathrm{out}}$. From this field, we directly compute beam intensities to form a continuous-space turbulence dataset. Simultaneously, we project $E_{\mathrm{out}}$ onto Laguerre-Gaussian (LG) modes indexed from $1$ to $\ell_{\max}$ and construct a turbulence-induced scattering matrix. This scattering matrix provides two complementary descriptions of the evolution of OAM. First, we interpret its columns as OAM-mode sorter outputs~\cite{oam_mode_sorter} and reconstruct OAM-basis qubits from the scattered modes, generating the discrete OAM-qubit intensity patterns shown in Fig.~\ref{fig: intro_fig}(a). Second, we use four pairs of modes $\pm \ell$ to construct effective single-qubit operator matrices that serve as turbulence channels. Propagating OAM-encoded qubits through these channels produces channel-dependent intensity patterns, which we combine with conventional quantum noise models to generate a channel-noise dataset. The resulting datasets establish a common platform for analyzing how channel dynamics generate structured leakage across both turbulence-derived and conventional quantum noise processes.

Having established the simulated channel dynamics, we next validate the resulting OAM evolution experimentally. We employ a 785~nm continuous-wave laser system to investigate continuous-space propagation and a bi-photon source to study discrete OAM-qubit evolution, as illustrated in Fig.~\ref{fig: intro_fig}(b). In both platforms, we introduce controlled turbulence by encoding Kolmogorov phase screens on a spatial light modulator (SLM). These experiments provide a direct connection between the simulated channel models and physically observed OAM propagation, allowing us to verify the predicted channel-induced distortions and leakage behavior across both classical and quantum optical regimes.

The simulated and experimentally measured datasets then serve as inputs to our distributed quantum feature-extraction framework as illustrated in Fig. \ref{fig: intro_fig}(c). Our central objective is to determine whether the structured leakage generated by channel evolution retains robust signatures beyond the ideal logical subspace. To accomplish this, we encode both ideal and degraded OAM intensity patterns into distributed quantum representations and process them using quantum convolutional operations (QConv). These operations map the original $M\times N$ intensity distributions into higher-dimensional feature spaces, where the dimensionality is determined by the number of qubits employed within each distributed quantum circuit. By partitioning the encoding task across smaller quantum subsystems, the distributed architecture reduces quantum-resource requirements while mitigating encoding noise. Using the framework described in Appendix~A, we measure expectation values in the $Z$ basis and construct feature maps that capture robust signatures arising from the underlying leakage dynamics.

\begin{figure*}[b!]
    \centering
    \includegraphics[width=.85\linewidth]{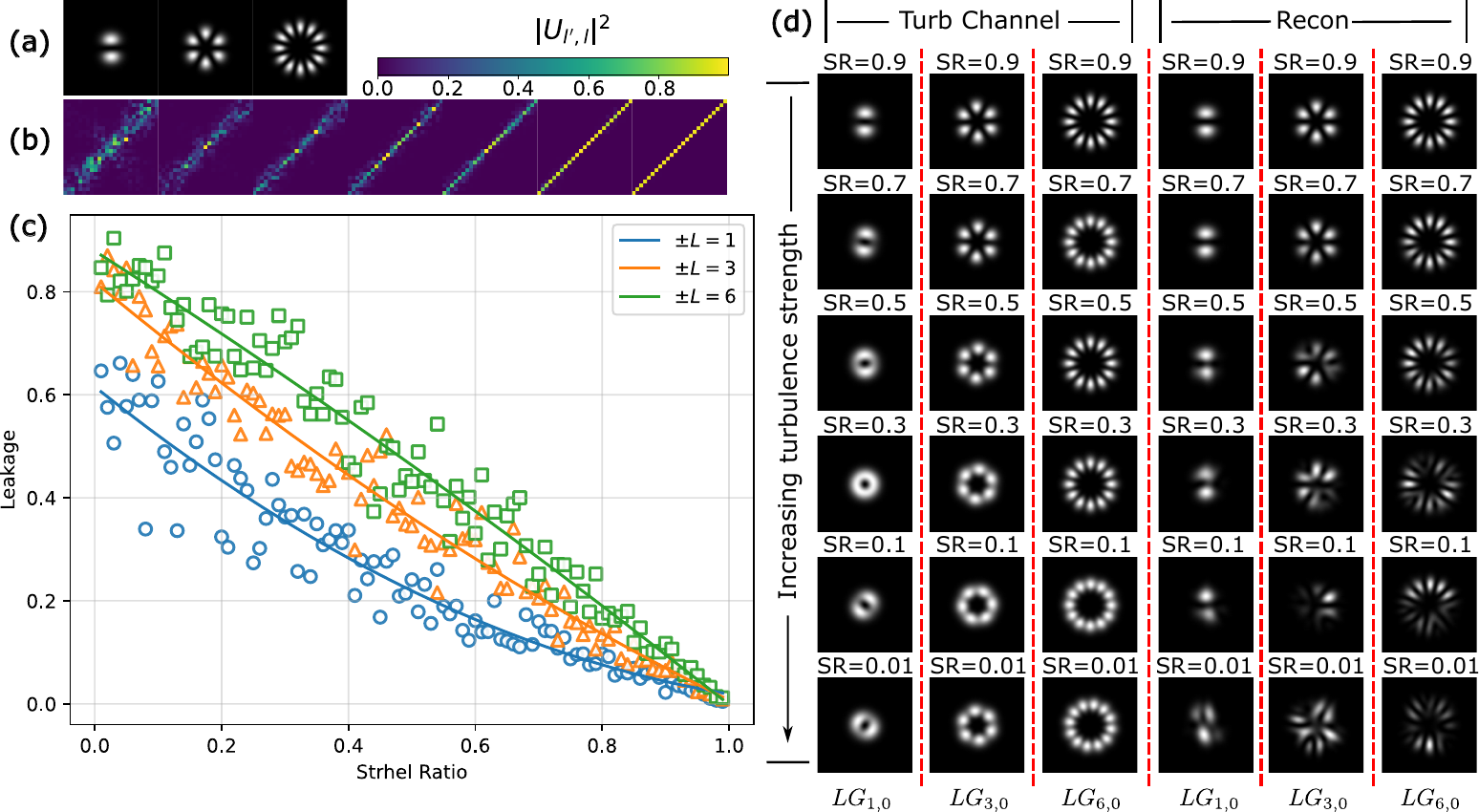}
    \caption{(a) Original intensity pattern with $\ell=\pm1, \pm3, \pm6$. (b) Scatter pattern varied in turbulence strength. From left to right, SR varies from 0.01 to 0.99. (c) Leakage trends as turbulence strength grows. Polynomial fitting is applied to the scatters. (d) Noisy pattern of turbulence channel and OAM-qubit reconstruction.}
    \label{fig: sr_leakage}
\end{figure*}

Finally, we evaluate propagation quality using the Strehl ratio (SR), which provides an optical measure of the severity of turbulence. We further employ the structural similarity index measure (SSIM) to compare feature maps and assess channel-induced distortions. To assess the simulated quantum evolution at the state level, we consider an idealized tomography framework, 
and use the reconstructed states to evaluate quantum-state fidelity. Together, these metrics enable us to quantify the relationship between channel dynamics, structured leakage, and high-dimensional feature representations, and to assess how effectively distributed quantum feature extraction preserves patterns across diverse channel conditions.

\subsection{Scatter pattern and leakage analysis}
First, we investigate how channel dynamics generate structured leakage in OAM qubits. Fig.~\ref{fig: sr_leakage} summarizes the relationship between turbulence strength, OAM-mode scattering, leakage, and OAM-qubit reconstruction. Together, these results establish the physical foundation for the distributed quantum feature-extraction framework developed in the following sections.

To model turbulence-induced channel dynamics, we initialize a fully superposed qubit state
\begin{equation}
    \ket{\psi} = \frac{1}{\sqrt{2}}\left(\ket{0}+\ket{1}\right), \ \rho =\ket{\psi}\bra{\psi}= \begin{pmatrix}
    \frac{1}{2} & \frac{1}{2} \\
    \frac{1}{2} & \frac{1}{2}
\end{pmatrix}.    
\end{equation}

To capture and reproduce the discrete implementation of the turbulence screen, we simulate the complete propagation process using the parameters summarized in Table~\ref{tab: sim_params}, which mimics the experimental configuration. We then project the propagated field onto the OAM basis to construct turbulence-induced scattering matrices, from which we derive both effective turbulence channels and reconstructed OAM-qubit states.

\begin{table}[h!]
    \centering
    \begin{tabular}{l l | l l}
        \hline \\
         Grid size (N $\times$ N) & 256 $\times$ 256 & Sample grid length ($L_x$) & $4 \times 10^{-3}$ m \\
         Wavelength ($\lambda$) & 785 nm & Gaussian beam waist ($\omega_0$) & $0.5 \times 10^{-3}$ m \\
         Wave-vector (k) & $2\pi/\lambda$ & OAM basis space ($\ell_{max}$) & 12 \\
         Distance between transmitter & 0.1 m & Number of screens ($n_{screens}$) & 5 \\
         and receiver ($z_{total}$) &&&\\
         OAM Index & varies from & Turbulence strength ($r_0$) & varies from  \\
         & $\ell = \pm 1$ to $\pm 6$ & & $4.071 \times 10^{-4}$ m to $1.608 \times 10^{-2}$ m \\
         &&&  (SR=0.01 to SR=0.99) \\
         \\
         \hline
    \end{tabular}
    \caption{Parameters for simulating the propagation process.}
    \label{tab: sim_params}
\end{table}

Fig.~\ref{fig: sr_leakage}(b) illustrates the evolution of the scattering matrix as the turbulence strength varies from destructive turbulence ($\mathrm{SR}=0.01$) to nearly turbulence-free propagation ($\mathrm{SR}=0.99$). Under weak or negligible turbulence, the scattering matrix is dominated by anti-diagonal components, indicating that the propagated field remains localized within its original OAM basis and experiences minimal crosstalk. As turbulence increases, scattering amplitudes spread away from the anti-diagonal structure and populate neighboring OAM modes. This redistribution reflects the emergence of structured leakage beyond the ideal logical subspace and provides a direct visualization of turbulence-induced channel dynamics. A detailed mathematical analysis of the scattering process is provided in Methods.

To quantify this behavior, we evaluate leakage across 100 randomly generated turbulence realizations spanning $\mathrm{SR}=0.01$ to $\mathrm{SR}=0.99$. As shown in Fig.~\ref{fig: sr_leakage}(c), leakage exhibits an approximately linear dependence on turbulence strength after polynomial fitting. We further observe that leakage increases systematically with OAM order, indicating that higher-order OAM states are increasingly susceptible to channel-induced degradation. Because the turbulence phase screen $\mathbb{M}_{NN}$ is stochastic, we additionally analyze 10,000 realizations in Appendix~B. These large-scale simulations confirm the persistence of the leakage trends and reveal clearly distinguishable leakage regions associated with different OAM orders.

The scattering matrix further enables us to construct effective turbulence channels and reconstruct OAM qubits within the logical basis. Representative examples are shown in Fig.~\ref{fig: sr_leakage}(d), where increasing turbulence leads to progressively distorted channel outputs and reconstructed OAM-qubit intensity patterns. The resulting channel-dependent intensity patterns provide a direct means of relating OAM-qubit channel dynamics to leakage-induced distortions in both continuous and discrete representations. More specifically, the simulated turbulent channel behavior shows a slight rotation along with loss of superposition. As the OAM order increases, the behavior becomes hard to identify through direct observation, which leads us to apply a distributed quantum convolutional operation for further evaluation.

\begin{figure*}[h!]
    \centering
    \includegraphics[width=.85\linewidth]{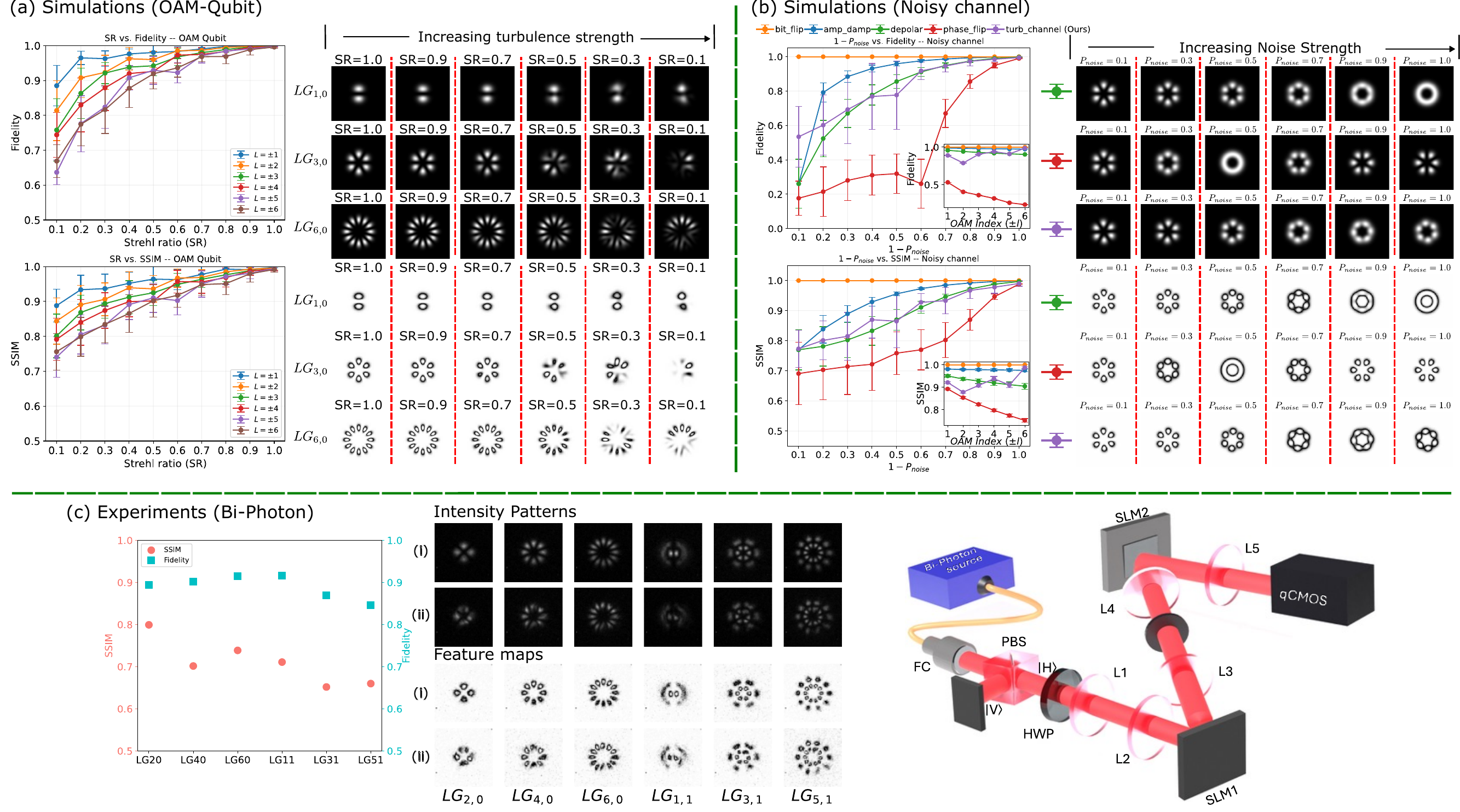}
    \caption{Discrete OAM-qubit and noisy channel behavior. (a) Simulation on OAM-qubit. (b) Simulation on noisy channel. Different color bars indicate simulated bit-flip (orange), amplitude damping (blue), depolarizing (green), phase-flip (red), and our turbulence channel (purple). For representative patterns, we focus on depolarizing (green), phase-flip (red) and our turbulence channel (purple), along with $\ell = \pm 3, \ p=0$. (c) Experiments with a bi-photon source. Sub-fig (a) and (b) provide the fidelity (upper-left) and SSIM (bottom-left) evaluations with a standard deviation error at each data point, along with the example pattern and extracted feature from $\ell = \pm1, \pm3, \pm6, \ \ p=0$. Sub-fig (c) provides the same two evaluations of $\ell=\pm2,\pm4, \pm6,  \ \ p=0 $ and $\ell=\pm1,\pm3,\pm5, \ \ p=1$, along with the pattern, extracted feature, and the optical experiment setup with a bi-photon source.}
    \label{fig: discrete_results}
\end{figure*}

\subsection{Discrete simulation and experimental behavior}

We now investigate how structured leakage-induced distortions manifest in reconstructed OAM states and how they differ from conventional quantum noise processes. Fig.~\ref{fig: discrete_results} summarizes the behavior of reconstructed OAM qubits under turbulence-derived and standard quantum channels. To analyze reconstructed OAM states in a unified quantum framework, we encode OAM intensity patterns into distributed quantum representations and process them using quantum convolutional operations, as illustrated in Fig.~\ref{fig: intro_fig}(c). The encoding maps OAM intensity patterns onto logical-qubit states for each OAM mode from $\ell = \pm1 \, \text{to} \pm6$, while the quantum convolutional layer transforms the resulting states into higher-dimensional feature representations designed to preserve channel-dependent features. For implementation, we employ rotation-$Y$ gates with a dynamic range of $L_{data}=8$, consistent with the unsigned 8-bit integer (Uint8) image-encoding standard. The convolutional layer is constructed using the 9-wire \textit{RandomLayer} \cite{randomlayer} circuit available in \textit{PennyLane}, which serves as a quantum analogue of a $3\times3$ convolutional kernel. To ensure reproducibility, we fix the \textit{RandomLayer} seed to 42 and generate nine circuit parameters using the \textit{NumPy} random number generator. Additional analyses of circuit initialization and parameter variance are provided in Appendix~C.
 
 \subsubsection{Simulation results} 
 We first examine the reconstructed OAM-qubit dataset generated from the turbulence scattering matrices introduced in the previous section. Representative examples are shown in Fig.~\ref{fig: discrete_results}(a). As shown in the fidelity and SSIM evaluation plot, with the decrease of SR, the reconstructed intensity patterns progressively deviate from their ideal OAM states. Moreover, we calculate the standard deviation due to the random factor $\mathbb{M}_{NN}$ and the nine copies from the output features. The error shrinks as SR increases in both fidelity and SSIM evaluation plots, indicating stable representation of OAM information at relatively high SR. We apply multiple OAM orders in this simulation, from $\ell = \pm 1$ to $\pm 6$, $p=0$, marked with varied colors. In the fidelity evaluation plot, as the order increases, the difference is distinct at lower SR (0.1 to 0.6) and minor at higher SR (0.7 to 1.0). Similar behavior is also observed in the SSIM evaluation plot, whereas the differences are smaller. In addition, overall trends are above 0.6 and 0.7 in the fidelity and SSIM evaluation plots, respectively. From the extracted feature, the contours become less distinct, the intensity patterns deform, and characteristic dragging behavior emerges within the reconstructed patterns. 
 
 Next, we compare turbulence-induced degradation with conventional quantum noise channels. In addition to turbulence-derived channels, we evaluate bit-flip (orange), phase-flip (red), depolarizing (green), and amplitude-damping (blue) channels together with the effective non-trace-preserving turbulence channels constructed from the scattering amplitudes. Representative results are shown in Fig.~\ref{fig: discrete_results}(b). To quantify the behavior, we apply the noise parameter $P_{\mathrm{noise}}$, and the general noisy-channel behavior is as expected. More specifically, the bit-flip channel has a minor effect on input states, because it mainly interchanges the amplitude of $\ket{0}$ and $\ket{1}$. The phase flip channel has an entirely opposite phase change at $P_{\mathrm{noise}} = 0.1$, and a loss of superposition at $P_{\mathrm{noise}} = 0.6$. However, the fidelity and SSIM trends for phase flip differ significantly because of the blank area after feature extraction, but we still observe a slight turning point between $P_{\mathrm{noise}} = 0.5$ and $0.6$. 
 
 We find similar trends for the amplitude damping and depolarizing channels, with consistently increasing fidelity and SSIM. However, the amplitude damping channel turns out to be more noise-resilient than the depolarizing channel. Furthermore, the turbulence channel shows a trend similar to that of the depolarizing channel, with minor differences under most of $P_{\mathrm{noise}}$, except $P_{\mathrm{noise}} = 0.1, 0.2, 0.5$ in the fidelity plot and $P_{\mathrm{noise}} = 0.4$ in the SSIM plot. The representative images also illustrate the corresponding behavior. The depolarizing channel gradually loses superposition, and the phase-flip channel destroys superposition in the middle and causes a certain degree of rotation along with reconstruction of superposition. 
 
 Furthermore, the turbulence channel turns out to be the combined behavior of amplitude damping and depolarizing channels. It contains slight rotation along with the destruction and reconstruction of superposition. We also fix the $1-P_{\mathrm{noise}}$ at 0.5, and simulate the trends under increasing OAM orders, as shown in the inset plots in both evaluation plots. As the OAM order increases, the conventional quantum channels turn out to have a consistent decreasing behavior, whereas the turbulence channel shows an unstable increase at $\ell = \pm2$ and $\pm5$. This behavior is also reflected in the main plots in Fig.~\ref{fig: discrete_results}(b). As we calculate the standard deviation of both fidelity and SSIM, the turbulence channel has a higher range of error than the conventional quantum channels. However, the overall errors of all simulated channels gradually decrease as SR increases. Note that, for conventional quantum channels, which are completely positive and trace preserving (CPTP), the dominant effect is a rotation of the reconstructed intensity pattern with relatively limited deformation of the underlying lobe structure. In contrast, turbulence-derived channels produce both rotational behavior and substantial deformation of the intensity pattern, consistent with leakage beyond the logical OAM subspace. 
 
 To be more specific for understanding the turbulence behavior, because the noise parameters associated with conventional quantum channels are not directly equivalent to the optical Strehl ratio, the apparent distortion scales differ across channel models. To better characterize turbulence-channel behavior, we generate ten independent realizations of the extracted turbulence operators and analyze the resulting ensemble response. As illustrated in Fig.~\ref{fig: discrete_results}(b) representative patterns, we observe that increasing channel strength initially produces both rotation and loss of superposition for $P_{\mathrm{noise}}=0.1$--$0.7$. As the noise level approaches $P_{\mathrm{noise}}=1.0$, rotational behavior remains evident while partial superposition recovery begins to emerge. These observations further distinguish turbulence-induced leakage from conventional quantum noise processes and demonstrate that turbulence produces structured channel-dependent distortions rather than simple unitary transformations.
 
\subsubsection{Experimental validation} 
To validate these observations experimentally, we perform turbulent OAM-qubit measurements using a spontaneous parametric down-conversion (SPDC) bi-photon source, illustrated in Fig.~\ref{fig: discrete_results}(c). The generated 810~nm photon pairs are coupled into a fiber coupler (FC) and separated using a polarizing beam splitter (PBS), where one polarization component is directed through a half-wave plate (HWP) toward SLM1. We then encode OAM states using holograms and propagate the photons through turbulence realizations generated on the SLM2. The resulting distorted OAM patterns are recorded using a single-photon-sensitive qCMOS camera (ORCA-Quest).

The limited photon flux results in low pixel intensities, while background noise further reduces the signal-to-noise ratio, particularly near $\mathrm{SR}=0.5$. To improve contour visibility during feature extraction, we employ a reduced encoding dynamic range of $L_{data}=5$. Despite these experimental limitations, the measured feature maps exhibit behavior consistent with the simulations, showing modest rotational effects accompanied by dominant turbulence-induced distortions. 

From the example patterns shown in Fig.~\ref{fig: discrete_results}(c), we select the pattern from $\ell = \pm 2, \pm 4, \pm6, \ p=0$, shown as the first 3 columns of Fig.~\ref{fig: discrete_results}(c) Intensity Patterns and Feature maps, to demonstrate consistent behavior with simulations. Then we further select $\ell = \pm 1, \pm 3, \pm5, \ p=1$, shown in the last 3 columns, to add an additional factor $p$ that might affect the QConv evaluation process. Row (i) indicates the original pattern without any turbulence, and row (ii) represents the distorted pattern after introducing turbulence at $\mathrm{SR} = 0.5$ to the original pattern. As the pattern shows a significant influence under classical Gaussian noise, our model still retains robust performance in detecting the contours, as shown in the feature maps. As for SSIM and fidelity evaluations, the values remain in the range above 0.6 and 0.8, respectively, which is relatively consistent with the extent of turbulence channel simulation results shown in Fig.~\ref{fig: discrete_results}(b). This consistent behavior may further reveal a potential relation between turbulence dynamics in quantum channels.

\subsection{Continuous-wave simulation and experimental behavior}

\begin{figure*}[h!]
    \centering
    \includegraphics[width=0.85\linewidth]{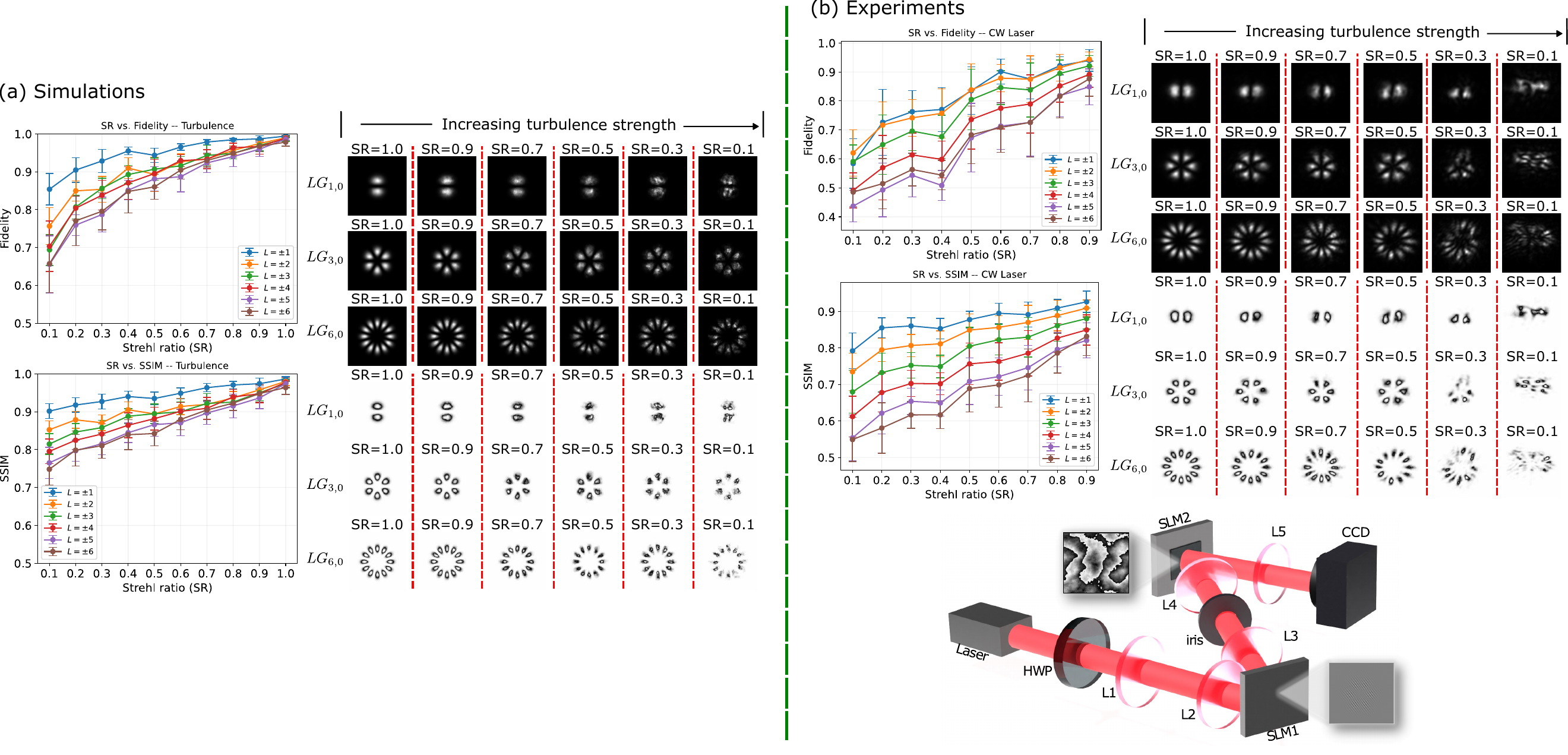}
    \caption{Continuous wave simulation and experimental results. (a) Simulation of continuous wave propagation. (b) Experiments with a 785nm CW-Laser. Sub-fig (a) and (b) provide the Fidelity (upper-left) and SSIM (bottom-left) evaluations, along with the example pattern and extracted feature from $\ell = \pm1, \pm3, \pm6, \ p=0$, along with the optical experiment setup with a continuous-wave laser source.}
    \label{fig: CW_sim_exper_plots}
\end{figure*}

To further examine the generality of the observed channel dynamics, we extend our analysis beyond the logical-qubit representation and investigate OAM propagation in the continuous domain. Fig.~\ref{fig: CW_sim_exper_plots} summarizes the corresponding simulation and experimental results. Our objective is to determine whether the leakage-induced distortions observed in reconstructed OAM qubits are also manifested in continuous OAM beams propagating through turbulence.

\subsubsection{Simulation results}

Using the propagation parameters listed in Table~\ref{tab: sim_params}, we simulate the evolution of continuous OAM beams through Kolmogorov turbulence and evaluate the resulting intensity distributions after propagation. The mathematical formulation of free-space propagation through turbulence is provided in Sections~4.2 -- 4.8. Representative examples are shown in Fig.~\ref{fig: CW_sim_exper_plots}(a). Consistent with the discrete OAM-qubit results, increasing turbulence strength progressively degrades the propagated intensity patterns. 

The fidelity and SSIM plots from the extracted features using QConv show a gradual rise with increasing SR and clear degradation with increasing OAM order, which are similar to the previous discrete results. As shown by the calculated standard deviation at each data point, the error decreases significantly in the fidelity evaluation plots, whereas the error decrease in the SSIM evaluation is relatively slight compared to the fidelity plots. In particular, as shown in the representative images, the characteristic OAM patterns become increasingly diffuse, with a gradual loss of contrast and spatial localization. As the Strehl ratio decreases, the intensity distribution transitions from a well-defined petal structure to a highly distorted profile, indicating increasing mode coupling and redistribution of optical energy across the OAM basis.

\subsubsection{Experimental validation}

We validate these observations experimentally using a 785~nm continuous-wave laser system, illustrated in Fig.~\ref{fig: CW_sim_exper_plots}(b). The laser beam is first expanded using lenses L1 and L2 before illuminating a spatial light modulator (SLM1, Santec SLM-200). Forked diffraction gratings displayed on SLM1 generate Laguerre--Gaussian superposition (petal) beams. The structured beam is subsequently relayed through a 4$f$ imaging system (L3 and L4), where an iris spatially filters the desired diffraction order. We emulate atmospheric turbulence by displaying Kolmogorov phase screens on SLM2, and record the resulting far-field intensity patterns using lens L5 and a CCD camera (Thorlabs CS2100M).

Representative measurements are shown in Fig.~\ref{fig: CW_sim_exper_plots}(b), where the initial state is prepared as $\ket{\phi} = \frac{1}{\sqrt{2}}(\ket{0}+i\ket{1})$. The evaluation plots show increasing trends consistent with the simulated results, starting from 0.4 in fidelity and 0.5 in SSIM. However, because of general randomness and noise from the natural environment, the error range changes only slightly from $SR=0.1$ to $0.7$ and decreases at $SR=0.8$ and $0.9$ in the fidelity evaluation plot. For SSIM, the error range remains nearly unchanged as SR increases. The corresponding behaviors are also illustrated in the representative pattern, as turbulence strength increases, the OAM intensity patterns not only fade but also begin to overlap and merge, producing distortions that closely resemble those observed in the reconstructed OAM-qubit states of Fig.~\ref{fig: discrete_results}(a). The close agreement between simulation and experiment demonstrates that the degradation mechanisms identified in the logical-qubit framework persist in continuous OAM propagation. These observations therefore provide complementary evidence that turbulence-induced channel evolution produces structured distortions across both discrete and continuous OAM representations.

\section{Discussion}

The central hypothesis of this work is that turbulence-induced degradation of OAM states is not purely stochastic but instead emerges from structured leakage generated during channel evolution. By representing OAM modes as logical qubits propagating through noisy quantum channels, we show that turbulence redistributes probability amplitude beyond the ideal logical subspace while preserving detectable signatures of the underlying channel evolution. The observed leakage trends and feature-space representations consistently support this interpretation. These results suggest that turbulence-induced degradation may be associated with structured channel dynamics that might be characterized through channel evolution and feature-space representations.

The logical-qubit representation provides a physically motivated connection between optical propagation and quantum-channel representations. Within this framework, the turbulence-induced scattering matrix serves not only as a description of mode crosstalk but also as a mechanism for constructing effective turbulence channels that govern OAM-qubit evolution. This perspective enables direct comparison between turbulence-derived channels and conventional quantum noise models, linking optical distortions with established quantum-channel dynamics. While conventional CPTP channels primarily induce transformations within the encoded logical subspace, turbulence channels generate mode-dependent crosstalk that redistributes probability amplitude beyond that subspace. The resulting evolution exhibits effective non-trace-preserving behavior within the truncated OAM basis and reveals structured leakage as a defining characteristic of turbulence-induced quantum noise.

Furthermore, a key outcome of this study is the identification of leakage as a physically meaningful indicator of channel-induced degradation. The approximately linear relationship between leakage and turbulence strength, together with its dependence on OAM order, suggests that leakage provides a quantitative measure of how probability amplitude is redistributed during propagation. Importantly, similar distortion behavior appears in both reconstructed OAM-qubit states and continuous-wave propagation experiments, indicating that the observed leakage is not specific to a particular representation. The consistency between discrete and continuous OAM propagation further suggests that structured leakage might be an intrinsic feature of OAM evolution through turbulent channels. 

Moreover, we have introduced a distributed quantum feature-extraction framework that operates directly on OAM-qubit representations. By transforming OAM states into higher-dimensional quantum feature spaces, the framework extracts robust features while reducing quantum-resource requirements through distributed encoding. The persistence of leakage-induced signatures across turbulence strengths, OAM orders, and propagation conditions suggests that physically meaningful structure may remain accessible even when direct inspection of the original intensity pattern becomes difficult. These findings indicate that turbulence-induced distortions preserve feature-space organization that can be exploited for characterization and robust feature extraction.

Several promising directions remain open for future investigation. More broadly, extending these concepts to larger, higher-dimensional, and more realistic scenarios could uncover new possibilities for robust characterization, information processing, and control within open quantum systems.

\section{Methods}
\subsection{Quantum Noisy Channel}
Quantum noisy channels are characterized as the operator-sum representation \cite{Kraus2}, 
\begin{equation}
    \begin{aligned}
        \varepsilon (\rho) &= \sum_k \bra{e_k} U \left[ \rho \otimes \ket{e_0} \bra{e_0} \right] U^{\dagger} \ket{e_k} \\
        &= \sum_kE_k \rho E_k^{\dagger},
    \end{aligned}
\end{equation}
where $\ket{e_0}$ is the initial state of the environment, and $\ket{e_k}$ denotes an orthonormal basis of the environment Hilbert space. By tracing over the environmental component after the joint unitary evolution $U$, the quantum channel can be expressed in the Kraus operator-sum representation, where the Kraus operators are defined as
\begin{equation}
E_k = \bra{e_k}U\ket{e_0}.
\end{equation}

The noisy models we explored in the noisy channel evaluations are listed below:
    \begin{equation}
        \begin{split}
            &\text{Bit-Flip:}\, E_0 = \sqrt{p}X, \ E_1 = \sqrt{1-p} I; \\
            &\text{Phase-Flip:}\, E_0 = \sqrt{p}Z, \ E_1 = \sqrt{1-p} I; \\
            &\text{Depolarizing:}\, E_0 = \sqrt{\frac{p}{4}}X, \ E_1 = \sqrt{\frac{p}{4}}Y, \ E_2 = \sqrt{\frac{p}{4}}Z, \ E_3 = \sqrt{1-\frac{3p}{4}}I; \\
            & \text{Amplitude Damping:}\,
             E_0 = 
            \begin{bmatrix}
            1 & 0 \\
            0 & \sqrt{1-p}
            \end{bmatrix}, \ E_1= 
            \begin{bmatrix}
                0 & \sqrt{p} \\
                0 & 0
            \end{bmatrix}.
        \end{split}
        \label{eq: noise_ele}
    \end{equation}
where $p$ represents the noise parameter marked as $P_{noise}$ in Section 2.3.1.

According to the Stinespring dilation theorem \cite{Stinespring1955}, any completely positive trace-preserving (CPTP) quantum channel can be represented as a unitary interaction between the system and an environment. For the special case of our theoretical turbulence channel, we will discuss this in Section 4.9.
\subsection{Transverse field representation}
All optical fields are represented on a square transverse plane 
$(x,y)\in[-L_x/2,L_x/2]^2$ sampled on $N\times N$ grid with spacing 
$\Delta x = L_x/N$. Cartesian coordinates are converted to polar form via $ r=\sqrt{x^2+y^2},\, \Phi=\mathrm{atan2}(y,x)$, which provides a natural basis for orbital-angular-momentum (OAM) modes.

\subsection{Laguerre--Gaussian OAM basis}
The input and output fields are expanded in the paraxial Laguerre--Gaussian
(LG) basis with radial index $p=0$ and azimuthal index $\ell$. Each mode is a logical basis state, and the pair $|\ell\rangle$ defines the logical qubit, which is given by
\begin{equation}
    |\ell\rangle \equiv \mathrm{LG}_{0}^{\ell}(r,\Phi)
    = \mathcal{N}_{\ell}
    \left(\frac{\sqrt{2}r}{w_0}\right)^{|\ell|}
    \exp\!\left(-\frac{r^2}{w_0^2}\right)
    e^{i\ell\Phi},
\label{eq:lg_beam}
\end{equation}
where $w_0$ is the beam waist and $\mathcal{N}_{\ell}$ ensures
$\langle \ell | \ell\rangle = 1$.
These modes form an orthonormal basis for the OAM subspace used in the
simulation.

\subsection{Free-space propagation: angular spectrum method}
Propagation over a distance $dz$ is modeled using the paraxial
angular-spectrum propagator. Spatial frequencies are defined as
\begin{equation}
    f_x = \frac{n-N/2}{L_x}, \qquad
f_y = \frac{m-N/2}{L_x},
\end{equation}
and the Fresnel transfer function \cite{goodman_intro_fourier_optics, david_computational_fourier_optics} is
\begin{equation}
    H(f_x,f_y;dz)
= \exp\!\left[-i\pi\lambda dz\left(f_x^2+f_y^2\right)\right].
\end{equation}
The propagated field is obtained via
\begin{equation}
    E(x,y;z+dz)
= \mathcal{F}^{-1}\!\left[
H(f_x,f_y;dz)\,
\mathcal{F}\{E(x,y;z)\}
\right],
\end{equation}
where $\mathcal{F}$ denotes the two-dimensional Fourier transform.

\subsection{Turbulence Simulation}
In this work, we applied a CW Laser directed onto a spatial-light modulator (SLM) with a given phase mask, $\Theta^{(\ell, -\ell)}$ to convert Gaussian mode optical beam to Laguerre-Gauss (LG) mode of $\pm \ell$, $u_{p,\ell}$, where $\ell$ indicates the degree of orbital angular momentum (OAM) that the mode contains. The turbulence with the strength quantified by the standard refractive index, $C_n^2$ is given by Eq. \ref{eq:1}.
\begin{equation}
    \Phi(X,Y) = \mathbb{R} \left\{ \mathcal{F}^{-1} \left(\mathbb{M}_{NN}\sqrt{\phi_{NN}(k)}\right)\right\},
    \label{eq:1}
\end{equation}
where $\phi_{NN}(k) = 0.023r_0^{-5/3}(k^2+k_0^2)^{-11/6}e^{-k^2/k_m^2}$ and the Fried's parameter $r_0 = (0.423k^2C_n^2d)^{-3/5}$. The mathematical symbol $\mathbb{R}$ represents the real part of the complex field and $\mathcal{F}^{-1}$ indicates the inverse Fourier transform operation. Furthermore, $k$, $d$, and $\mathbb{M}_{NN}$ denote the wave number ($2\pi/\lambda$), the propagation distance, and the encoded random matrix, respectively. In the previous results section, our OAM-qubit and turbulence simulation apply randomly generated $\mathbb{M}_{NN}$ to mimic the natural environment. We provide Appendix~D for fixed $\mathbb{M}_{NN}$ to avoid any possible effects caused by this. Instead of using traditional $C_n^2$ to represent the turbulence strength, we apply Strehl Ratio in the experiments, which is given by Eq. \ref{eq:2}.

\begin{equation}
    S \approx exp(-\sigma^2)
    \label{eq:2}
\end{equation}
where $\sigma^2$ refers to the wavefront variance calculated as $\sigma^2=1.03(D/r_0)^{5/3}$ according to Noll's relationship between the wavefront variance and the ratio $D/r_0$.

Under this context, people have proposed an assumption that in a single-photon LG mode with radial index $p=0$, the quantum state evolution of turbulent fluctuation can be described by a unitary operator $U_{turb}$, which transforms the input state $|\ell_0\rangle$ into a superposition of many OAM modes \cite{unitary_turb_relation},
\begin{equation}
    \ket{\psi_{\ell_0}} = U_{turb}\ket{\ell_0} = \sum_{\ell} c_{\ell, \ell_0} \ket{\ell},
\end{equation}
with $c_{\ell,\ell_0}=\langle \ell|U_{turb} \ket{\ell_0}$ that relates the Kraus operators in quantum channel operation with the turbulence in a pure quantum state.

\subsection{Kolmogorov Phase Screen Normalization}

When the turbulence is modeled using $n_{\text{screens}}$ independent
phase screens, the variances add:
\begin{equation}
    \sigma_{\phi,\text{total}}^2
    = n_{\text{screens}}\,\sigma_{\phi,\text{screen}}^2.
\end{equation}
Equating this to the Kolmogorov prediction gives the per-screen RMS:
\begin{equation}
    \sigma_{\phi,\text{screen}}
    = \frac{\sqrt{1.03}}{\sqrt{n_{\text{screens}}}}
      \left( \frac{D}{r_0} \right)^{5/6}.
\end{equation}
Let $\phi_{\text{raw}}(x,y)$ be the unnormalized phase generated from the
Kolmogorov PSD \cite{fast_sim_kolmogorov}, and let $\sigma_{\text{sim}} = \mathrm{RMS}\!\left[ \phi_{\text{raw}}(x,y) \right]$ computed over the beam region. We scale the phase by a factor $\alpha$, $\phi(x,y) = \alpha\,\phi_{\text{raw}}(x,y)$, so that $\mathrm{RMS}[\phi] = \sigma_{\phi,\text{screen}}$.  
Since RMS scales linearly with $\alpha$, the normalization factor is
\begin{equation}
    \alpha
    = \frac{\sigma_{\phi,\text{screen}}}{\sigma_{\text{sim}}}
    = \frac{1}{\sigma_{\text{sim}}}
      \cdot
      \frac{\sqrt{1.03}}{\sqrt{n_{\text{screens}}}}
      \left( \frac{D}{r_0} \right)^{5/6}.
\end{equation}
This ensures that each simulated phase screen has the correct Kolmogorov
statistics for the chosen $r_0$, beam diameter $D$, and number of
screens $n_{\text{screens}}$.

\subsection{Split-step propagation through turbulence}
propagation through turbulence over a total distance $z$ is
approximated using a split-step method with $n$ phase screens:

\begin{equation}
   E(z+dz)
\approx
e^{\frac{i}{2k}\nabla_\perp^2 dz/2}
\, e^{i\phi(x,y)} \,
e^{\frac{i}{2k}\nabla_\perp^2 dz/2}
E(z), 
\end{equation}
where $dz=z/n$. This operator sequence provides a numerical solution to the paraxial wave equation with refractive-index fluctuations.

\subsection{Parameter Consistency and Validity}

We consider the following parameters: $w_0 = 0.5\times 10^{-3}\,\text{m}, \, z_{\text{total}} = 0.1\,\text{m}, \, n_{\text{screens}} = 5, \, r_0 = 1.5\times 10^{-3}\,\text{m},\, L_x = 4\times 10^{-3}\,\text{m},\,
N = 256,\, \lambda = 785\times 10^{-9}\,\text{m}.$

\subsubsection{Beam size and turbulence strength}

The beam diameter is $D = 2w_0$, and per-screen RMS phase is $\sigma_{\phi,\text{screen}}
    = \frac{\sigma_\phi}{\sqrt{n_{\text{screens}}}} \approx 0.73$ radians.

\subsubsection{Spatial sampling: beam and $r_0$ resolution}

The transverse grid spacing is $ \Delta x = \frac{L_x}{N} = 15.6~\mu\text{m}.$ The beam waist in pixels is $\frac{w_0}{\Delta x} \approx 32$, so the beam diameter spans $\sim 64$ pixels, well within the $256\times256$ grid. Similarly, the Fried parameter in pixels is $\frac{r_0}{\Delta x} \approx 96$. Thus, the turbulence correlation length $r_0$ is sampled by $\sim 96$ pixels, which is more than sufficient to resolve the Kolmogorov structure.

\subsubsection{Fresnel propagation sampling in $z$}

For FFT-based Fresnel propagation on a square grid of size $L_x$ with
sampling $\Delta x$, a commonly used sampling condition for the
propagation step $\Delta z$ is
\begin{equation}
    \Delta z = \frac{z_{\text{total}}}{n_{\text{screens}}} = 0.02\, \text{m}  \;\lesssim\; \Delta z_{\max}
    = \frac{N\,\Delta x^2}{\lambda} = 0.079\, \text{m}, 
\end{equation}
so each Fresnel step is safely within the sampling limit, and aliasing from under-sampling in $z$ is not expected.

\subsection{Turbulent scattering matrix}
For each input OAM mode $|\ell_{in}\rangle$, the propagated
field $E_{\mathrm{out}}^{(\ell_{\mathrm{in}})}$ is projected onto the OAM basis to obtain the scattering matrix given by,
\begin{equation}
    U_{\ell_{\mathrm{out}},\ell_{\mathrm{in}}}
=
\langle \mathrm{LG}_{0}^{\ell_{\mathrm{out}}}
\mid
E_{\mathrm{out}}^{(\ell_{\mathrm{in}})}
\rangle.
\end{equation}

The matrix $|U|^2$ quantifies turbulence-induced OAM crosstalk. In the full infinite dimensional Hilber space, the evolution operator is unitary. However, numerical simulations truncate the OAM basis to $|\ell\rangle \in [-|\ell_{max}\rangle, \, |\ell_{max}\rangle]$. Turbulence scatters amplitude into arbitrarily other logical basis, and the projection onto the truncated basis yields a sub-unitary matrix,
\begin{equation}
    U = PUP,\, U^\dagger U \neq I,
\end{equation}
where $P$ is the projector onto the simulated OAM basis. The deviation from unitarity quantifies the probability lost to modes outside the truncated basis.

A logical qubit is encoded in the two-dimensional subspace, $|0\rangle = |\ell\rangle$, $|1\rangle = |-\ell\rangle$. The logical qubit experiences an effective quantum channel obtained by extracting the $2\times 2$ block
\begin{equation}
    U_q =
\begin{pmatrix}
U_{+\ell,+\ell} & U_{+\ell,-\ell} \\
U_{-\ell,+\ell} & U_{-\ell,-\ell}
\end{pmatrix}.
\end{equation}
Each turbulence realization yields a Kraus-like operator $U_q^{k}$, and the ensemble-averaged quantum evolution is
\begin{equation}
\rho_{\mathrm{out}}
=
\frac{1}{N}
\sum_{k=1}^{N}
U_q^{k}\,
\rho_{\mathrm{in}}\,
U_q^{k\dagger}.
\end{equation}
Because $U_q^{k\dagger}U_q^k < I$, the map is completely positive but not trace-preserving, reflecting physical leakage of probability amplitude into other OAM bases outside the logical subspace. A CPTP channel can be recovered by introducing an erasure operator as,
\begin{equation}
    K_{loss} = \sqrt{I - \sum_k U_q^{k\dagger}U_q^k }.
\end{equation}

\subsection{Spatial intensity of the logical density matrix}
Given the output density matrix

\begin{equation}
    \rho =
\begin{pmatrix}
\rho_{00} & \rho_{01} \\
\rho_{10} & \rho_{11}
\end{pmatrix},
\end{equation}
the corresponding spatial intensity distribution is
\begin{equation}
    I_\rho(x,y)
=
\rho_{00}|\langle\ell|\ell\rangle|^2
+
\rho_{11}|\langle-\ell|-\ell\rangle|^2
+
\rho_{01}\langle-\ell|\ell\rangle
+
\rho_{10}\langle\ell|-\ell\rangle.
\end{equation}

\subsection{Quantum Convolutional Operation}
For an image $X=[x_{i,j}]_{H\times W}$, for an $K\times K$ convolutional kernel, we define the patch as $P_{u,v} = [x_{u+i, v+j}]^{K-1}_{i,j=0}$. A two-dimensional discrete classical convolution operation could be defined as \cite{2d_conv_paper}
\begin{equation}
    (X * W)_{u,v}=\sum^{K-1}_{i=0}\sum^{K-1}_{j=0} x_{u+i, v+j}w_{i,j},
\end{equation}
where $W$ is the convolution kernel, and $u,v$ are the starting position of the patch. In the quantum convolutional operation, we start from quantum angle encoding. For a single patch, we have
\begin{equation}
    \begin{aligned}
        \ket{\psi_{u,v}^{(0)}}&=\otimes^{K-1}_{i,j=0}\left( \cos\frac{\phi_{i,j}}{2}\ket{0}+\sin\frac{\phi_{i,j}}{2}\ket{1} \right), \\
        \phi_{i,j}&=\frac{\pi x_{u+i, v+j}}{2^{L_{data}}-1}\in[0,\pi],
    \end{aligned}
\end{equation}
where $L_{data}$ represents the dynamic range of the input data. The further Random-Layer operation could be expressed as a parameterized unitary operation of $n_{param}$ parameters $\theta=(\theta_0, \theta_1, \cdots, \theta_{n_{param}-1})$,
\begin{equation}
    \ket{\psi_{u,v}} = U(\theta)=\ket{\psi_{u,v}^{(0)}}
\end{equation}
where it will create $m\times n$ copies of the convoluted values. To retrieve the output feature, we calculate the $k^{th}$ Z-basis measurement expectation value $\langle z_k(u,v)\rangle$ from $m \times n$ copies for each sliding window iteration\cite{randomlayer, vqa},
\begin{equation}
    \langle z_k(u,v)\rangle = \bra{\psi_{u,v}^{(0)}} U^{\dagger}(\theta)Z_kU(\theta)\ket{\psi_{u,v}^{(0)}}.
\end{equation}
We pick one consistent measurement output from each iteration. After post-processing to calculate the absolute value, normalize to UINT8 format, and rearrange, we will get the output features $F=[z_k(u,v)]_{H' \times W'}$. Due to the sliding window function, the dimension changes follow the same classical convolution equation,
\begin{equation}
    \begin{aligned}
        H'&=\left\lfloor \frac{H+2P-K}{S} \right\rfloor + 1 \\
        W'&=\left\lfloor \frac{W+2P-K}{S} \right\rfloor + 1
    \end{aligned}
\end{equation}
where $P$, $S$ are padding and stride, respectively.

The quantum convolutional layer we use is the random layer, which randomly chooses single-qubit rotations and 2-qubit entanglement gates to generate the parameterized quantum circuit (PQC), also known as an Ansatz in quantum machine learning.

\subsection{OAM decomposition and leakage analysis}
The full propagated field is decomposed into the OAM basis. The logical qubits of the field and basis are represented as,

\begin{equation}
    \rho_{out} = |E_{out}\rangle\langle E_{out}|,\quad
    \hat{O_\ell} =  \{|\ell\rangle\langle \ell|\}, \quad 
    c_{\ell} = Tr(\rho_{out}\hat{O}), \quad
    \rho_{recon} = \sum_{\ell=-\ell_{\max}}^{\ell_{\max}} c_\ell |\ell\rangle\langle \ell| = |E_{recon}\rangle\langle E_{recon}|.
\end{equation}

The normalized leakage outside the logical subspace is quantified by
\begin{equation}
    I_{leakage} = 1 -
\frac{|c_{+\ell}|^2 + |c_{-\ell}|^2}
{\sum_{\ell}|c_\ell|^2}.
\end{equation}

\subsection{Combined Mean Fidelity}
As the state scale grows, especially in a huge information space, fidelity decay \cite{fidelity_decay} occurs and drags the value extent to a relatively low scale, so we apply combined mean fidelity instead. Assume we have two quantum states $\ket{\psi}=\ket{\psi_0}\otimes\ket{\psi_1}\otimes\cdots\otimes \ket{\psi_{N-1}}=\otimes^{N-1}_{i=0}\ket{\psi_i}$ and $\ket{\phi}=\ket{\phi_0}\otimes\ket{\phi_1}\otimes\cdots\otimes \ket{\phi_{N-1}}=\otimes^{N-1}_{i=0}\ket{\phi_i}$, we have
\begin{equation*}
    \begin{aligned}
        \bar{F}(\psi, \phi) &= \sqrt[N]{|\bra{\psi}\phi\rangle|^2} = |\otimes^{N-1}_{i=0}\bra{\psi_i} \otimes^{N-1}_{i=0}\ket{\phi_i}|^{2/N} \\ 
        &= \left| \left(\bra{\psi_0}\otimes \bra{\psi_1} \otimes\cdots\otimes\bra{\psi_{N-1}}\right) \left( \ket{\phi_0} \otimes \ket{\phi_1}\otimes \cdots \otimes\ket{\phi_{N-1}} \right) \right|^{2/N} \\
        &= \left| \bra{\psi_0} \phi_0\rangle \otimes \bra{\psi_1}\phi_1\rangle \otimes \cdots \otimes\bra{\psi_{N-1}}\phi_{N-1}\rangle \right|^{2/N} \\
        &= \left|\prod^{N-1}_{i=0} \bra{\psi_i}\phi_i\rangle \right|^{2/N} = \prod^{N-1}_{i=0}|\bra{\psi_i}\phi_i\rangle|^{2/N}=\sqrt[N]{\prod^{N-1}_{i=0}F(\psi_i, \phi_i)}
    \end{aligned}.
\end{equation*}
Due to the calculation complexity, we apply the natural log instead
\begin{equation*}
    \begin{aligned}
        \bar{F}(\psi, \phi) = \sqrt[N]{\prod^{N-1}_{i=0}F(\psi_i,\phi_i)} = \exp\left(\frac{1}{N}\sum^{N-1}_{i=0} \ln \left(F(\psi_i, \phi_i)\right)\right)
    \end{aligned}.
\end{equation*}

In the context of a noisy channel, it is often represented in density matrix format
\begin{equation*}
    \begin{aligned}
        \bar{F}(\rho, \sigma) = \sqrt[N]{\prod^{N-1}_{i=0}\left(\operatorname{Tr}\sqrt{\sqrt{\rho_i}\sigma_i \sqrt{\rho_i}} \right)^2}  = \exp\left(\frac{1}{N}\sum^{N-1}_{i=0} \ln \left(F(\rho_i, \sigma_i)\right)\right)
    \end{aligned}.
\end{equation*}

\subsection{Structural Similarity Index Measurement}
In order to statistically analyze how the distributed circuit mitigates channel noise in embedded images, we use the structural similarity index measure (SSIM). The SSIM is defined as

\begin{equation}
    SSIM(F_{\rho}, F_{\rho'}) = \frac{(2\mu_{F_{\rho}}\mu_{F_{\rho'}} + c_1)(2\sigma_{F_{\rho},F_{\rho'}} + c_2)}{(\mu_{F_{\rho}}^2 + \mu_{F_{\rho'}}^2 + c_1)(\sigma_{F_{\rho}}^2 + \sigma_{F_{\rho'}}^2 + c_2)},
\end{equation}
where $\mu$ denotes the mean intensity of each OAM pattern. The terms $\sigma^2_{F_{\rho}}$ and $\sigma^2_{F_{\rho'}}$ represent the sample variances calculated from individual windows of reference and noise signals, respectively. The sample covariance between these windows is denoted by $\sigma_{F_{\rho}, F_{\rho'}}$. To ensure numerical stability when the denominator is small, the constants $c_1$ and $c_2$ are introduced. These constants are functions of the dynamic range of the noisy intensity $L_{noise}=\max(F_{\rho'})-\min(F_{\rho'})$. Specifically, $c_1$ and $c_2$ are defined as
\begin{equation}
    c_1 = (k_1 L_{noise})^2, \quad c_2 = (k_2 L_{noise})^2,
\end{equation}
where $k_1 = 0.01$ and $k_2 = 0.03$ are two default constant coefficients used in general image processing.

\bibliography{sample}

\section*{Acknowledgments}

S.L. acknowledges support from the SMU Startup Fund and the Sam Taylor Fellowship Program. The experimental work at UIC was supported by the U. S. Department of Energy (DE-SC0024096 and the Quantum Internet to Accelerate Scientific Discovery Program under Field Work Proposal No. 3ERKJ381). The work performed by N.R. was supported in part by the Gordon and Betty Moore Foundation through Grant GBMF13927. 

\section*{Author contributions statement}

Y.M. developed the quantum convolutional framework, performed the simulations, and generated the figures. Y.M. and S.L. developed the software. N.R. and T.S. conducted the experiments. Y.M., S.S., and S.L. analyzed the results. All authors contributed to the manuscript and approved the final version. S.L. conceived and led the project.

\section*{Competing interests} 
The authors have no conflicts of interest to declare.
\section*{Appendix A. Distributed Quantum Computation}
Due to the tremendous grid space we have in the OAM intensity pattern and the expansion of the quantum convolutional filter, we apply a distributed quantum system \cite{yayu_2025qce_workshop_distri_comp_noisy_channel, yayu_2025qce_poster_distri_comp} for accelerating the computation process.
\begin{figure*}[h!]
    \centering
    \includegraphics[width=1.0\linewidth]{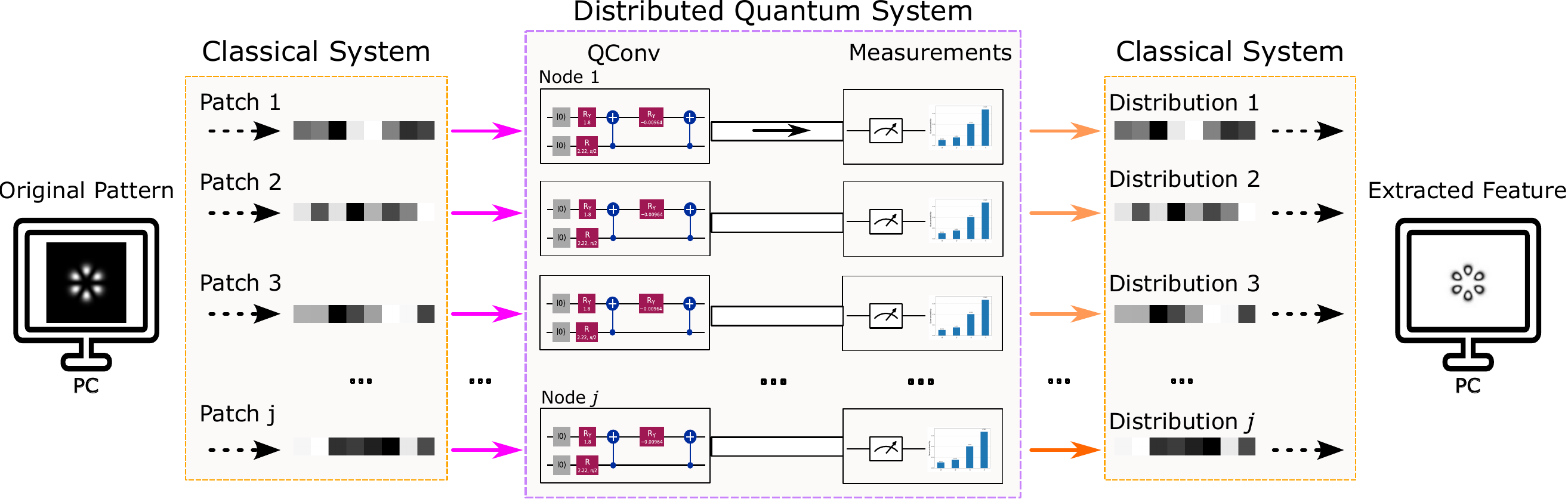}
    \caption{Distributed Quantum System}
    \label{fig: dis_q_sys}
\end{figure*}

As shown in Fig. \ref{fig: dis_q_sys}, the basic idea of this distributed system is to divide the patches into a classical system and feed them separately into quantum computation nodes. Then collect the measurement results and combine them for evaluation. In our previous work \cite{yayu_2025qce_workshop_distri_comp_noisy_channel, yayu_2025qce_poster_distri_comp, yayu_2026dcas_distri_img_encoding}, we demonstrate the potential of eliminating the noise by increasing the node numbers. This potential could also be applied in this work to control the unnecessary noise generated by circuits.

\section*{Appendix B. Sampling of normalized leakage under random $\mathbb{M}_{NN}$}
As discussed in the previous section, due to the random nature of the turbulence phase screen factor $\mathbb{M}_{NN}$, we calculate polynomial fits from 100 leakage samples. In this appendix, we aim to demonstrate the trends and diversity under varied $\ell$s with no further fitting operation.

With the support of the NVIDIA L4 GPU, we are able to simulate over 10,000 random $\mathbb{M}_{NN}$ samples varying from $SR=0.0001 \approx 0$ (destructive turbulence) to $SR=0.9999 \approx 1$ (no turbulence) and $\ell = \pm1$ to $\ell = \pm6$.

\begin{figure*}[h!]
    \centering
    \includegraphics[width=0.5\linewidth]{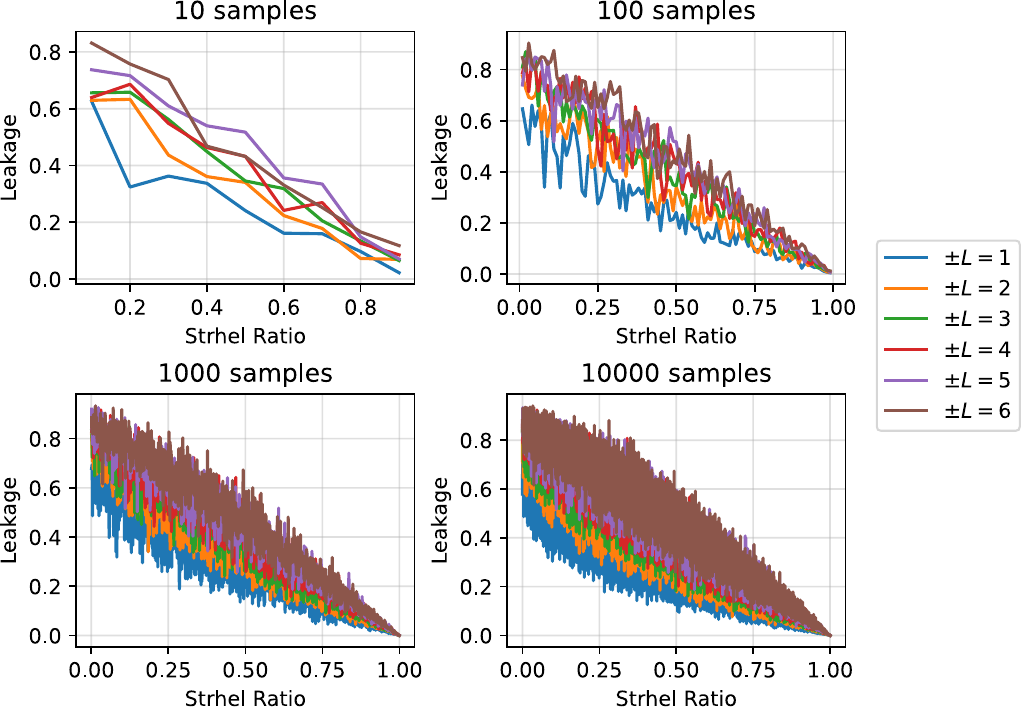}
    \caption{Samples 10, 100, 1000, 10000}
    \label{fig:samples}
\end{figure*}

As shown in Fig. \ref{fig:samples}, from 10 to 1000 samples, the trends are consistent with Fig. \ref{fig: sr_leakage} (c). The distinguishable color area from $\ell=\pm1$ to $\ell=\pm6$ also shows that as $\ell$ increases, the leakage gets severe.

\section*{Appendix C. Consistency under varied Random-Layer seed and parameter size}
\begin{figure*}[h!]
    \centering
    \includegraphics[width=0.5\linewidth]{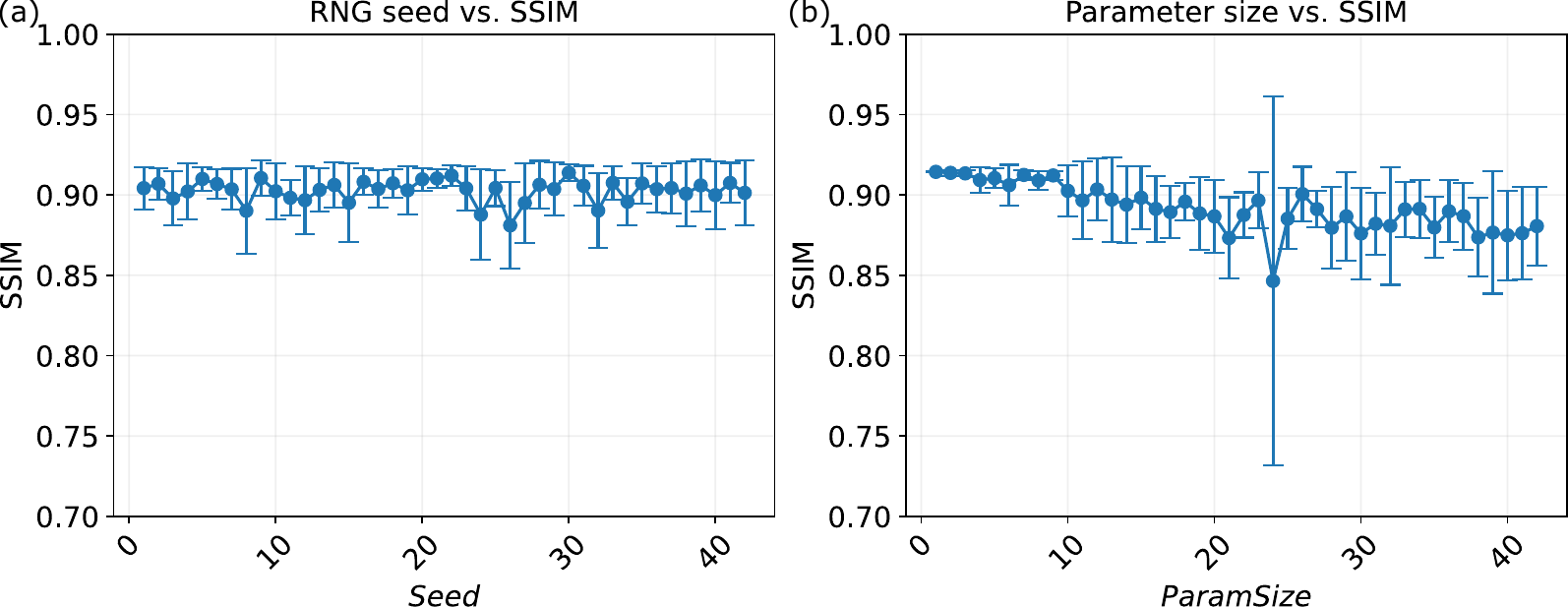}
    \caption{(a) PennyLane random number generator (RNG) seed consistency. We set the parameter of this evaluation to 9 to fit the evaluation in the main text. (b) Parameter size consistency. We set the seed of this evaluation to 42 to fit the evaluation in the main text.}
    \label{fig: seed_param_size_SSIM}
\end{figure*}

As we discussed in the previous section, the evaluation quantum convolutional circuit, \textit{RandomLayer}, contains two main random properties: (1) a seed to generate a random circuit. (2) the size of random parameters that affect the combination of gates. Therefore, we add two sets of evaluations to eliminate the influence of randomness. As shown in Fig. \ref{fig: seed_param_size_SSIM}, it turns out that the RNG seed and parameter size have a minor effect on the feature extraction process. 

\section*{Appendix D. Theoretical behavior with fixed $\mathbb{M}_{NN}$}
\begin{figure*}[h!]
    \centering
    \includegraphics[width=1.0\linewidth]{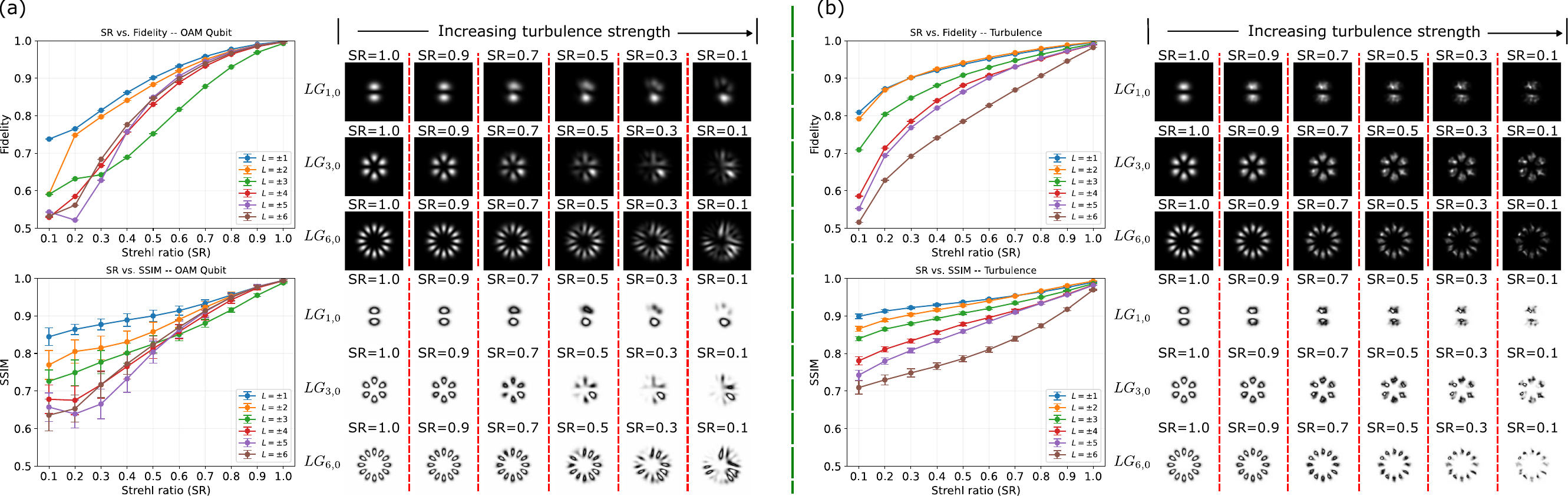}
    \caption{(a) OAM-qubit pattern with fixed $\mathbb{M}_{NN}$ (b) Turbulence pattern with fixed $\mathbb{M}_{NN}$}
    \label{fig:sim_oam_qubit_fixed_cn}
\end{figure*}

In the previous results and methods section, our OAM-qubit and turbulence simulation apply a randomly generated $\mathbb{M}_{NN}$ to mimic the natural environment. From the results, the curves appear to bounce slightly, and the fading of OAM intensity pattern is not following a consistent pattern. Therefore, we add two more evaluations in fixed $\mathbb{M}_{NN}$, for which we set a seed in each iteration of data generation. The results, as shown in Fig. \ref{fig:sim_oam_qubit_fixed_cn}, turn out to be smoother, and the OAM intensity lobe changes follow a consistent pattern.

\end{document}